\documentstyle[aas2pp4]{article}

\lefthead{Novosyadlyj \& Chornij}
\righthead{Quasar abundance as a probe of initial spectrum}

\begin{document}

\def\solar{\ifmmode_{\mathord\odot}\else$_{\mathord\odot}$\fi}

\title{QUASAR ABUNDANCE AT HIGH REDSHIFTS AS A PROBE
\\
 OF INITIAL POWER SPECTRUM ON SMALL SCALE}

\author{B. Novosyadlyj, Yu. Chornij}
\affil{ Astronomical Observatory of L'viv State University,
 Kyrylo i Methodij str.8, 290005 L'viv, Ukraine}

\begin{abstract}
The dependence of the number density of the bright QSOs at different
redshifts ($n_{QSO}(z)$) on initial power spectrum is studied. It is assumed
that QSO phenomenon is an early short term stage of evolution of massive
galaxies with $M\geq 2\times 10^{11}h^{-1}M_{\odot}$. The duration of
such QSO stage which is passed through by fraction $\alpha$ of galaxies
is determined by means of minimization of the divergence of the
theoretical number density of QSOs at different redshifts for specified
initial spectrum from observable one [\cite{sc91}]. It is shown that the
nearest number densities of QSOs at $0.7\le z\le 3.5$ to observable ones are
obtained for the tilted CDM model ($\Omega_{b}=0.1$, $n=0.7$).
The QSO stage lasts $\sim 7\times 10^{7}/\alpha$ years and begins soon
after the moment of rise of the first counterflow in collisionless
component and shock wave in gas.

The possibility of the reconstruction of initial power
spectrum on small scale on the base of the observable data on number density
of QSOs at different $z$ is considered too. Such reconstructed spectrum
in comparison with standard CDM has steep reducing of power at
$k\ge 0.5 h Mpc^{-1}$.
\end{abstract}

\keywords{cosmology:initial power spectrum - dark matter - quasars}

\section{Introduction}

 The problem of formation of galaxies and large scale structure of
the Universe has an astrophysical aspect consisting in search of
connection between properties of luminous objects
and dark matter halo, as well as cosmological aspect consisting in
determination of the initial power spectrum of density fluctuations.
Obtaining such spectrum from first principals would essentially
simplify the solving of problem on the whole.
Unfortunately, today we can not surely say which
the postinflation spectrum is: scale invariant $P_{pi}=Ak^{n}$ with
$n=1$, tilted with $n\not=1$, or more complicated. The nature
of dark matter (baryonic, collisionless cold dark matter, futher CDM,
hot dark matter, futher  HDM, or their mixture H+CDM, or other), and
values of cosmological parameters $h \equiv {H_{0} \over 100km/s\,Mpc}$
($H_{0}$- Hubble constant), $\Omega_{t} \equiv {8 \pi G
\over 3H_{0}^2}\rho_{t}$ ($\rho_{t}$ - total mean density),
$\Omega_{b} \equiv {8 \pi G \over 3H_{0}^2}\rho _{b}$ ($\rho_{b}$ - mean
density of baryons) and cosmological constant $\Lambda$ remain uncertain.
More preferable from theoretical point of view spectra (standard CDM
model, tilted CDM, CDM+$\Lambda$ or hybrid H+CDM) give predictions
consistent with the observable data marginally only [\cite{dav92}-\cite{hn95}].
Futhermore, the different spectra normalised to the COBE r.m.s. cosmic
microwave background (CMB) anisotropy predict close characteristics of
the large scale structure of the Universe, e.g. $\Delta T/T$ at degree
angular scale, large scale peculiar velocity field etc.
[\cite{pog93,nov94,hn95}] so that it is
impossible to distinguish them. Therefore, it is very
important to build up the new cosmological tests  based on extension of a
class of objects with predicted observational manifestations sensitive
to initial power spectrum on small scale. Here we shall investigate the
dependence of the number density of quasars (QSOs) at different redshifts
against initial power spectra. To make it possible we suppose that a QSO
is the early short term stage of evolution of a massive galaxy and appears
in the corresponding peaks of initial random Gaussian field of density
fluctuations. The number density of already collapsed peaks of certain
scale at different redshifts $z$ strongly depends on an amplitude and
slope of the power spectrum of such field at small scales
($k\sim 0.1-10h\,Mpc^{-1}$).

 Such approach was applied independently by some authors. For example,
Cen et al. [\cite{cen92}] have calculated the comoving number density of nonlinear
objects as a function of redshift z using Press-Schechter approximation
[\cite{pr}]. Their results show that observable number density of
quasar-like objects between $2\le z\le 4$ can be explained by tilted
CDM with $n\ge 0.7$ and mass of host galaxies $\le 5\;10^{12}M_{\odot}$.
Blanchard et al. [\cite{bl93}] tested HDM models and have shown the disability
of these
models to explain the observable  concentration of QSOs at $z\sim 3-4$
when primordial scale-invariant spectrum ($n=1$) is normalised to the
COBE r.m.s. cosmic  microwave background anisotropy [\cite{sm92}].
Kashlisky [\cite{ka93}] tested the hybrid H+CDM and modified CDM models by
number density of distant QSOs at $z=4.5$. The number density of
the peaks-precursors of massive galaxies ($M>10^{11} h^{-1}M_{\odot}$),
which passed a quasars stage soon after their central regions had
collapsed, was calculated for the different spectra. The mean number
density of QSOs predicted by modified CDM model with
$\Omega h=0.1,\;0.2,\;0.3$ and H+CDM was found to be less than observable
one $n_{QSO}^{obs}\approx 400 h^3Gpc^{-3}$ at $z=4.5$. Nusser \& Silk 
[\cite{nu93}]
for finding $n_{QSO}(z)$ in redshift range 2-5 for given spectrum used
the Gaussian peaks formalism by Bardeen et al. [\cite{b86}] and shown that this
formalism is more preferable than Press-Schechter one. They assumed
the number density of very bright QSOs ($L\ge 10^{47} erg/s$) to be equal
to the number density of
peaks with masses $M\ge 2\times 10^{11}h^{-1}M_{\odot}$ within certain height
interval corresponding to the physically substantiated duration of quasar
stage ($\approx 4\times 10^{7}$ years). They have concluded that tilted CDM
model with $n=0.8$, H+CDM ($\Omega_{CDM} \approx
0.7,\; \Omega_{HDM} \approx 0.3$) and scaled CDM with $0.2\leq \Omega h\leq 0.3$
are compatible with the mean abundance of quasars at $2\leq z\leq 5$.
Ma \& Bertchinger [\cite{ma}] have carried out the numerical simulation of
formation of elements of large scale structure of the Universe with
masses $10^{11}-10^{13}M_{\odot}$ in hybrid H+CDM models and analysed the
possibility of explanation of the observable number density of QSOs
at $2.2\leq z \leq 4.5$ in the framework of the Press-Schechter formalism.
They have shown that H+CDM model with $\Omega_{HDM}=0.2$ predicts
$n_{QSO}(z)$ which matches the  observable one if quasars are created soon
after collapse of nearly spherical peaks.

  The goal of the paper is an improvement of this approach to the explanation
of the observable data on abundance of the bright QSOs in wider range
$z=0.7 \div 4.7$ [\cite{sc91,bo88}], testing of the preferable initial
spectra, which marginally match other observable data on large scale
structure of the Universe, search of spectrum momenta to reproduce
the observable data on  $n_{QSO}(z)$ and construction the
suitable phenomenological spectrum on small scale.

\section{The main assumptions and method}

  We explore the distribution of quasar abundance over redshift $n_{QSO}(z)$
in the framework of the theory of formation of the large scale structure of
the Universe in the peaks of random Gaussian field of scalar density
fluctuations as consequence of gravitational instability on the Einstein-de
Sitter cosmological background. In this approach galaxies  and objects of
other scales are formed in  the peaks of density fluctuations which have an
amplitude $\delta \equiv {\delta \rho \over \rho} \ge \nu_{th}\sigma_{0}(R_{f})$,
where  $\sigma_{0}$ - r.m.s. amplitude  of the fluctuations of certain
scale  $R_{f}$, $\nu_{th}$ - some threshold height which corresponds to
minimal amplitude when objects of this scale forms still.

    In the theory of random Gaussian fields [\cite{b86,dor70}] a comoving number
density of the peaks with height  $\nu \geq \nu_{th}$ is

$$n(\nu_{th})=\int_{0}^{\infty}\;p(\nu;\nu_{th},q)\;N_{pk}(\nu)\;d\nu,\eqno(1)$$
where $p(\nu;\nu_{th},q)={(\nu/\nu_{th})^{q} \over 1 +
(\nu/\nu_{th})^{q}}$- threshold function, $N_{pk}(\nu)$ is
differential number density of peaks in the amplitude range  ($\nu, \nu+d\nu$),
 which depends on $\nu$ and spectrum momenta
  $\sigma_{0},\;\sigma_{1},\;\sigma_{2}$ (formulae 4.3-4.5 and A15 in the
  paper by Bardeen et al. [\cite{b86}]).

       We suppose that certain fraction of massive galaxies $\alpha_{1}$
passes through quasar stage of evolution. Maybe, only some fraction of them
$\alpha_{2}$ is visible by Earth observer. Therefore, the fraction of galaxies
passing through quasars stage and visible by Earth observer as  QSOs is
$\alpha \equiv \alpha_{1}\alpha_{2} \le 1$. If we adopt the black hole model of
quasars, then the minimum mass associated with a host galaxy is $\sim
10^{11}-10^{12}M_{\odot}$ [\cite{nu93,tu91,lo94}]
For selection of galactic scale peaks with  mass   $M \geq 2\times
10^{11} h^{-1}M_{\odot}$ we smoothed the initial random Gaussian density
field $\delta(\overrightarrow{x})$ by the Gaussian filter function -
$\overline{\delta}(\overrightarrow{x} ;R_{f})=
{1 \over (2\pi R_{f}^{2})^{1.5}}
\int \delta(\overrightarrow{x'})\;e^{-{(\overrightarrow{x'}-
\overrightarrow{x})^{2} \over 2\pi R_{f}^{2}}} d^{3}\overrightarrow{x'}$
with $R_{f}=0.35 h^{-1}$Mpc. It means that the power spectrum of the smoothed
field is $P(k;R_{f})=P(k)e^{-k^{2}R_{f}^{2}}$, where
$P(k) \equiv <\delta _{k}\delta^{*}_{k}>$ is initial power spectrum, $\delta _{k}$ is
Fourier transformation of the  $\delta(\overrightarrow{x})$:
$\delta(\overrightarrow{x})=
{1 \over (2\pi)^{1.5}} \int \delta_{k} {e^{i\overrightarrow{k}\overrightarrow{x}}} d^{3}\overrightarrow{k}$.
 Those among them which have an amplitude
$\overline{\delta}(\overrightarrow{x},R_{f})
\geq \nu^{g}_{th} \sigma_{0}$, where $\sigma_{0}^{2}=
{1 \over (2\pi)^{2}} \int dk\;k^{2}\;P(k)\;e^{-k^{2} R_{f}^{2}}$ is r.m.s.
amplitude of density fluctuations, evolve to massive galaxies. Their
concentration is known from $C_{f}A$ cataloque [\cite{dav82}]
$n_{g}^{obs}\approx (4.6h^{-1}Mpc)^{-3}$. Using eq.(1) for massive galaxies
$$n(\nu_{th}^{g},q^{g})=n^{obs}_{g}\eqno(2)$$
one parameter of threshold function  (e.g. $\nu_{th}^{g}$) can be found.

   As follows from Tolmen model, a homogeneous spherical-symmetrical
adiabatic fluctuation on the Einstein-de Sitter cosmological background
collapses at
$$z_{c}\equiv \left({t_{0} \over t_{c}}\right)^{2 \over 3}-1=0.59\sigma_{0}\nu_{c}-1,\eqno(3)$$
where $t_{0} \simeq 1.3 \times 10^{10}$ years is present cosmological time (here and further
we assume $H_{0}=50 {km \over s\;Mpc}$), $t_{c}$ - the moment of collapse,
$\nu_{c}$ - corresponding  height of peak. At this moment $t_{c}$
appearence of the first counterflows in dark matter and generation of shock
wave in baryon component take place. Massive black hole in central region of
such peak may be formed during $\sim10^{8}$ years as it follows from the
physical models of quasar mechanism (see, for example, [\cite{lo94}]).
Therefore, we suppose that QSO stage of galaxy evolution begins soon after
collapse of central region of peak $t_{c}$ or later by certain time interval
$\Delta t$ called a time delay. Duration of the QSO stage or
quasar lifetime is $\tau_{QSO}$.

        In such approach the number density of the QSOs at redshift $z$
is the number density of corresponding peaks of density fluctuations
(precursors of massive galaxies) which had collapsed between
$t_{c}-\tau_{QSO}$ and $t_{c}$. These peaks we can see as QSOs at redshift
$z$ corresponding to time $t(z)=t_{c}(z_{c})+ \Delta t$:
$$z=\left({t_{0} \over t_{c}(z_{c})+\Delta t}\right)^{2 \over 3}-1.\eqno(4)$$

    The height of such peaks $\nu$ is in the range of
($\nu_{c},\nu_{c}+\Delta \nu$), where
$$\nu_{c}=1.69\sigma_{0}^{-1}(z_{c}+1),\eqno(5)$$
$$\Delta\nu=1.13(z_{c}+1)^{2.5} {\tau_{QSO} \over \sigma_{0}t_{0}},\eqno(6)$$
as it follows from eq. (3). Then the number density of QSOs is
$$n_{QSO}(z)=\alpha\int_{\nu_{c}}^{\nu_{c}+\Delta\nu}\;
p(\nu;\nu_{th},q)\;N_{pk}(\nu)\;d\nu\eqno(7)$$
where $z$, $\nu_{c}$ and $\Delta\nu$ are defined by eq. 4, 5, 6 respectively. As we
can see, the number density of QSOs $n_{QSO}(z)$ depends  upon the duration
of the quasar stage $\tau_{QSO}$, the time delay $\Delta t$, parameters of
the threshold function $\nu_{th}$ and $q$, momenta of spectrum smoothed by
galactic filter $\sigma_{0}$, $\sigma_{1}$, $\sigma_{2}$
and fraction $\alpha$ of peaks-precursors passing quasar stage and
visible by Earth observer. If quasar stage is long-term
($\tau_{QSO}\sim t_{0}$) then the number density of QSOs at all $z$ is equal to
fraction of already collapsed peaks-precursors of massive galaxies
visible from Earth at quasar stage of their
evolution: $n_{QSO}(z)=\alpha n_{g}(z)$. If quasar stage is short-term
($\tau_{QSO}\ll t_{c}$), then it follows from (7) and (6) that
   $$n_{QSO}(z)=1.13\alpha \; {(z_{c}+1)^{2.5} \over \sigma_{0}t_{0}} \;
   p(\nu;\nu_{th},q)\;N_{pk}(\nu)\;\tau_{QSO}.\eqno(8)$$
It is valid for above mentioned estimation of $\tau_{QSO}\sim 10^{7}-10^{8}$ years.

        We suppose that the duration of the quasar stage $\tau_{QSO}$, the
time delay $\Delta t$ and parameters of threshold function $\nu_{th}$ and $q$
do not depend on the $z$. Also we assume that the sample of bright quasars given
by Schmidt et al. [\cite{sc91}] and Boyle et al. [\cite{bo88}] is complete for all $z$ in the
range 0.7-4.7. Here we do not discuss the problem concerned with the observational
data but pay particular attention to the theoretical approach of their using
for cosmological problems.

         Now we can calculate the number density of QSOs for set of
redshifts and given values of the duration of QSO stage $\tau_{QSO}$, the
time delay $\Delta t$, the fraction $\alpha$ of massive galaxies passing
quasar stage, parameters $\nu_{th}$ and q of threshold function and momenta of
spectrum $\sigma_{0}$, $\sigma_{1}$, $\sigma_{2}$. For present spectrum
normalised in a certain way the momenta $\sigma_{0}$, $\sigma_{1}$, $\sigma_{2}$ are
calculated independently. The time delay $\Delta t$ for given spectrum we
find in following manner. Using eq.(8) we calculate the redshift of maximum
$z_{max}$ of dependence  $n_{QSO}(z)$. If  $z_{max}$ is higher than one of
observable $n^{obs}_{QSO}(z)$  [\cite{sc91}] - $z_{max}^{obs}\approx 2.2$
then we find the time delay $\Delta t$ from equation
$$\left.{dn_{QSO}(z) \over dz}\right| _{z_{max}^{obs}=2.2} = 0.\eqno(9)$$
In those models where $z_{max}$  is less than or equal  $z_{max}^{obs}\approx 2.2$
we assume $\Delta t=0$. Rest four values $\alpha$, $\tau_{QSO}$, $\nu_{th}$
and $q$ are unknown. Finding of them by method of
minimization of divergence vector is a matter of testing of known spectra.

We try also to reconstruct the initial power spectrum from observable data
on number density of QSOs in wide range of redshifts and massive galaxies
on $z\approx 0$. For this we consider momenta of spectrum
$\sigma_{0}$, $\sigma_{1}$, $\sigma_{2}$ as unknown values and find them
in the same way.

         In all cases we write eq.(7) for 6 points of data by Schmidt et al.
[\cite{sc91}] on the $z_{1}=4.7$, $z_{2}=4.05$, $z_{3}=3.7$, $z_{4}=3$, $z_{5}=2.8$,
$z_{6}=2.2$, for 3 points on curve by Boyle et al. [\cite{bo88}] on $z_{7}=2.0$,
$z_{8}=1.5$, $z_{9}=0.7$ and the eq. (2) for number density of massive galaxies
on $z_{10}=0$:
$$\alpha\int_{\nu_{i}}^{\nu_{i}+\Delta\nu_{i}}\;p(\nu;\nu_{th},q)\;N_{pk}(\nu)\;
d\nu=n^{obs}_{qso}(z_{i}),\;(i=1,2...9),\eqno(10)$$
$$\int_{0}^{\infty}\;p(\nu;\nu_{th},q)\;N_{pk}(\nu)\;d\nu =n^{obs}_{g},\eqno(11)$$
 where $\nu_{i}$, $\Delta \nu_{i}$ are calculated for   $z_{i}$ accordingly to
(5) and (6) respectively. So, we solve the overdetermined system of
equations (10), (11) (or (10) only) by method of minimization of the divergence
vector using MathCad [\cite{mc}] software.

\section{Results}

\subsection{Testing of spectra}

        We shall test the preferable initial power spectra of density
fluctuations on the flat Friedmanian background expected in standard CDM
model ($\Omega_{CDM}=0.9$, $\Omega_{b}=0.1$), tilted CDM ones and hybrid
H+CDM one ($\Omega_{CDM}=0.6$, $\Omega_{HDM}=0.3$, $\Omega_{b}=0.1$), which
marginally match other observational data on large scale structure of
the Universe [\cite{dav92} - \cite{hn95}].
The  transfer function $T(k)$ in the spectra $P(k)=Ak^{n}T^{2}(k)$,
where $A$ is constant of normalisation
\footnote{The parameter $t_{1}$ of transfer function fitting formula by
Holtzman [\cite{ho89}] are accepted equal to 1.},
is taken from the paper by
Holtzman [\cite{ho89}].  All spectra we normalised to produce the COBE data on
CMB temperature anisotropy at angular scales
$10^{o}$ [\cite{ben94,wr94}]:
$<({\Delta T \over T})^{2}>^{1 \over 2}=(1.1\pm 0.1)\times 10^{-5}$.
We calculate $<({\Delta T \over T})^{2}>=C(0;10^{o})$
using the approximation formula by Wilson
and Silk [\cite{ws81}] for correlation function of CMB temperature anisotropy
measured by receiver with Gaussian response function and exact formula
[\cite{mgs89,nov96}] for correlation function
$C(\alpha)\equiv <{\Delta T \over T}(0)\;{\Delta T \over T}(\alpha)>$.
Both Sachs-Wolfe and Doppler
effects are taken into account (see also [\cite{nov96}]).
The contribution of tensor mode to $\Delta T/T$ is different in different
models of inflation. In the most of them it is not dominating at the COBE
angular scale (see review by [\cite{wh94}] and references cited therein),
in the some ones it is negligible even if $n<1$ (for example "natural"
inflation, [\cite{ad92}]). In this work, like ones cited here, we
consider only models of inflation which give no significant tensor mode.
But one can easily cross over to models of inflation with significant
tensor mode when ratio $\alpha_{T}\equiv \left(\Delta T/T\right)_{T}/
\left(\Delta T/T\right)_{S}$ is known (here $\left(\Delta T/T\right)_{T}$
and $\left(\Delta T/T\right)_{S}$ is contribution of tensor and scalar
modes correspondingly). The constant of normalization then will be
$A'=A/(1+\alpha_{T}^{2})$, the momenta of spectrum
$\sigma_{j} '=\sigma_{j}/\sqrt{1+\alpha_{T}^{2}}$, peaks-precursors of QSOs
will be collapsed at $z_{c}'=(z_{c}+1)/\sqrt{1+\alpha_{T}^{2}}-1$, where
values without ($'$) are ones calculated without tensor mode.
 Parameters of
cosmological models, constants of normalisation $A$ of power spectra,
their momenta $\sigma_{0}$, $\sigma_{1}$, $\sigma_{2}$ and biasing parameters
$b \equiv 1/\sigma_{8}$, where
$\sigma_{8}$ is the r.m.s.  mass fluctuations in the top-hat sphere with
radius $8h^{-1}Mpc$, are presented in Tabl.1 . (For comparison, our
normalization constant A for CDM model agrees with analogous constant from
paper [\cite{bun95}] reduced to h=0.5 and $\sigma(10^{o})=30.5\mu K$ with
accuracy better then 0.1\%).

\begin{table*}[tbh]
\caption{
The constant of normalisation of the spectra, theirs momenta
$\sigma_{0}$, $\sigma_{1}$, $\sigma_{2}$ on galactic scale
($R_{f}=0.35h^{-1}Mpc$) and biasing factor $b$, defined by
$b=\sigma_{8}^{-1}$, where $\sigma_{8}$ is $r.m.s.$ mass fluctuation 
on a scale of $8h^{-1}Mpc$.
}
\vspace{0.5cm}
\begin{center}
\begin{tabular}{|c|c|c|c|c|}
\hline
&\multicolumn{3}{|c|}{CDM}&$H\;+\;CDM$ \\
\hline
$n$           & 1            &   0.8        &0.7           &1             \\
\hline
$A$           &$7.33\times 10^{6}$&$1.62\times 10^{6}$&$7.57\times 10^{5}$&$7.51\times 10^{6}$ \\
$\sigma_{0}$  &   4.78       &   2.37       &   1.67       &  1.39        \\
$\sigma_{1}$  &   4.77       &   2.18       &   1.47       &  0.91        \\
$\sigma_{2}$  &   7.93       &   3.50       &   2.31       &  1.32        \\
$b$           &   1.05       &   1.77       &   2.30       &  1.66        \\
\hline
\end{tabular}
\end{center}
\end{table*}

For beginning we analyse the dependence of the $n_{QSO}(z)$
on $\tau _{QSO}$, $\Delta t$, $\nu _{th}$ and $q$ for different spectra
accepting  $\alpha$ equal to 1. Results are presented in
\mbox{Fig.1-3.}. 

\begin{figure*}[tbh]
\epsfxsize=16truecm
\centerline{\epsfbox{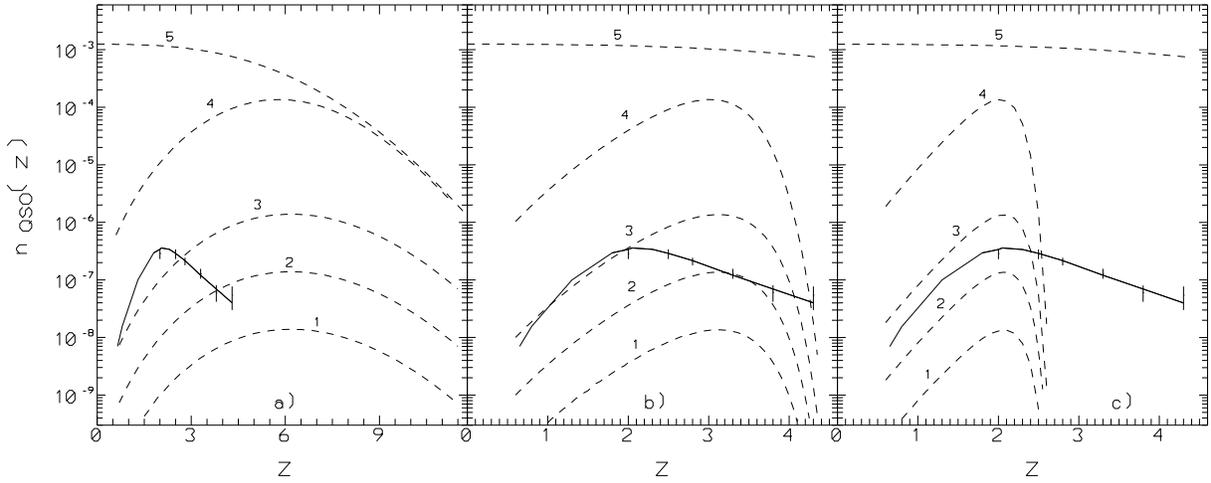}}
\caption{Quasar abundance against redshift
for different values of the
$\tau_{QSO}=10^4$, $10^5$, $10^6$, $10^8$, $10^10$ years (dashed lines
1, 2, 3, 4, 5 respectively in each panel) and time delay $\Delta t=0$
(a), $\Delta t =10^9$ years (b), $\Delta t=1.61\times 10^9$ years (c)
for CDM spectrum with $n=1,\;\Omega_{CDM}=0.9$ and $\Omega_b=0.1.$
The solid line with error bars is the comoving space density of quasars
from Schmidt et al. (1991). The upper dashed line (5) is the number density
of already collapsed peaks-precursors of bright galaxies. In all cases
$\alpha =1$.
}
\end{figure*}

\begin{figure*}[tbh]
\epsfxsize=16truecm
\centerline{\epsfbox{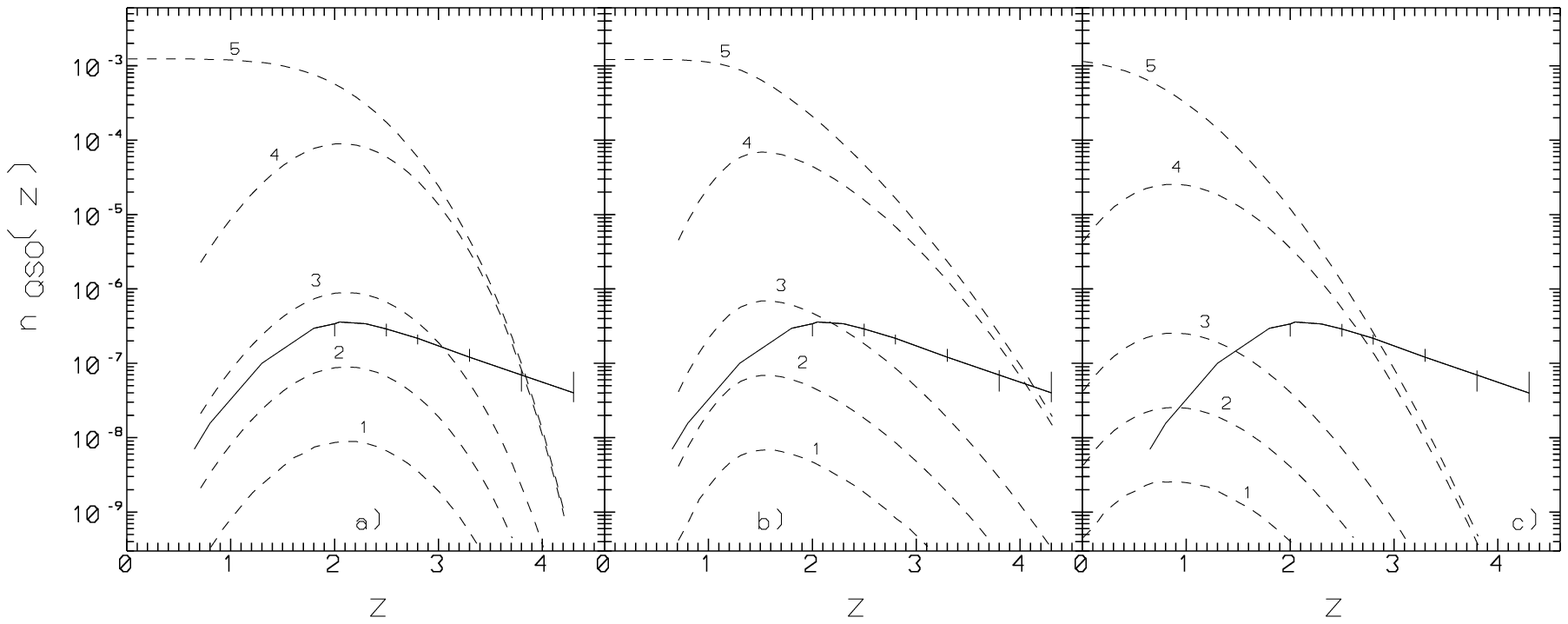}}
\caption{
Quasar abundance against redshift for CDM spectrum with
$\Omega_{CDM}=0.9$, $\Omega_{b}=0.1$, $n=0.8$ (a), $n=0.7$ (b)
and hybrid H+CDM model ($\Omega_{HDM}=0.3$, $\Omega_{CDM}=0.6$,
$\Omega_{b}=0.1$) for different values of the
$\tau_{QSO}=10^{4}$, $10^{5}$, $10^{6}$, $10^{8}$, $10^{10}$ years (dashed lines
1, 2, 3, 4, 5 respectively in each panel). Time delay $\Delta t$
is $5.2\times 10^{8}$ years (a), 0 (b) and (c). The rest is the same as in
Fig. 1.
}
\end{figure*}

As we can see, the number density of QSOs is proportional to
$\tau_{QSO}$, meanwhile the time delay displaces the maximum of $n_{QSO}$
to lower $z$ (Fig.1). Reducing of power spectrum at small scale results in
steeper declination of $n_{QSO}(z)$ at high $z$ and displacing of its
maximum to lower $z$ (Fig.1-2).
 Increasing of the value of the threshold parameter $q$ ($\tau_{QSO}$,
$\nu_{th}$ are constants) reduces the number density of QSOs at low
redshift and displaces its maximum to high $z$ (Fig. 3a).
Dependence $n_{QSO}(z)$ on $\nu_{th}$ ($\tau_{QSO}$ and $q$ are constants) is
similar (Fig. 3b). Increasing of $q$ with finding of corresponding $\nu_{th}$,
which ensures $n_{g}=n_{g}^{obs}$, results in steeper slope of $n_{QSO}(z)$
at low redshifts $z \le 2$ (\mbox{Fig. 3c}).
The curve labeled by 5 in Fig. 3 is the number density of QSOs calculated
without threshold function ($\nu_{th}=0$). For comparison we have calculated
also the $n_{QSO}(z)$ without threshold function for another minimal linear
overdensity corresponding to a bound object at $z=0$ $\delta^{0}_{c}=1.33$
adopted by Nusser and Silk [\cite{nu93}] (line labeled by 6 in Fig. 3a).

  Now let us find for each model such values of $\alpha,\; \tau_{QSO}$,
 $\nu_{th}$ and $q$ which give minimal divergence $n_{QSO}(z)$
from observable one and ensure the
number density of massive galaxies at $z\approx 0$. In all cases analyzed here
the amplitude of theoretical $n_{QSO}(z)$ is roughly equal to observable
$n^{obs}_{QSO}(2.2) \simeq 3 \times 10^{7}\;h^{3}Mpc^{-3}$ when
$\tau_{QSO} \sim 10^{6}$ years that is $\ll t_{0}$. It means that (8)
is good approximation of (10) in the range $10^{-3} \le \alpha \le 1$ and
that $n_{QSO} \propto \alpha \tau_{QSO}$. As the result, we can not
find both values $\alpha$ and $\tau_{QSO}$ simultaneously but their product
$\alpha \tau_{QSO}$ only.
Approximate solutions of system of equations (10)-(11) are obtained by using
of MathCad software package and presented in the Table 2. 
The quasar abundances $n_{QSO}(z)$ for them are shown in the Fig.4. 

\begin{table*}[tbh]
\caption{
Solutions $\alpha\tau_{QSO}$, $\nu_{th}$ and $q$ of system of
equations (10)-(11)  for different spectra.
}
\vspace{0.5cm}
\begin{center}
\begin{tabular}{|c|c|c|c|c|}
\hline
&\multicolumn{3}{|c|}{CDM}&$H\;+\;CDM$ \\
\hline
$n$           & 1            &   0.8        &0.7           &1             \\
\hline
$\alpha\tau_{QSO}$(years)& $3.54\times 10^{5}$ & $4.56\times 10^{5}$ & $2.39\times 10^{5} $ & $3.82\times 10^{4}$ \\
$\Delta t$ (years) & $1.55\times 10^{9}$ & $5.19\times 10^{8}$ &   0.00          &  0.00         \\
$\nu_{th}$         &   5.22         &   2.73         &   2.52          &  2.37        \\
$q$                &   1.95         &   6.60         &   12.4          &  4.30        \\
\hline
\end{tabular}
\end{center}
\end{table*}

\begin{figure*}[t]
\epsfxsize=16truecm
\centerline{\epsfbox{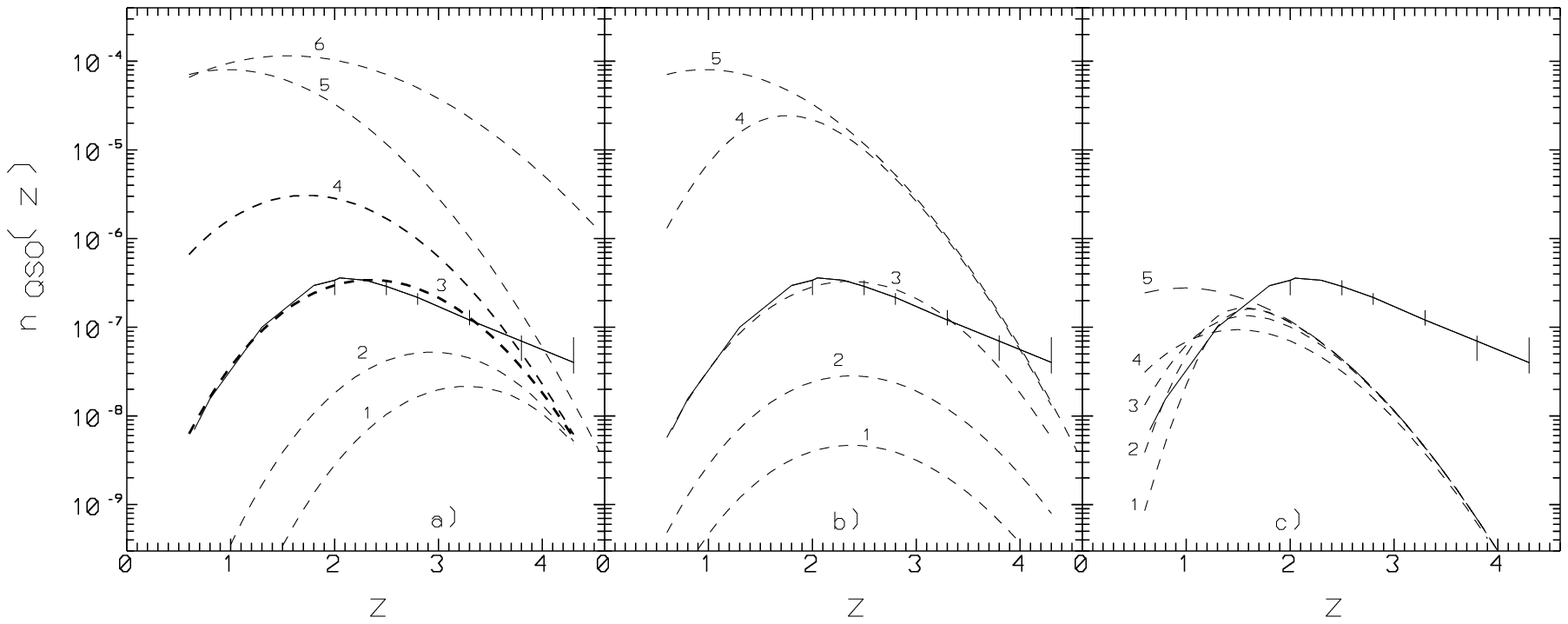}}
\caption{
Quasar abundance against redshift in tilted CDM model ($n=0.7$)
for different parameters of threshold function:
(a) - $q=16$ (line 1), $q=12$ (2), $q=8$ (3), $q=4$ (4), without
threshold function (5),
$\tau_{QSO}=6.9\times 10^{7}$ years, $\nu_{th}=5.89$;
(b) - $\nu_{th}=10$ (line 1), $\nu_{th}=8$ (2), $\nu_{th}=5.89$ (3),
$\nu_{th}=3$ (4), without threshold function (5),
$\tau_{QSO}=6.9\times 10^{7}$ years, $q=8.08$;
(c) - $q=16,\;\nu_{th}=2.5$ (line 1), $q=12.4,\;\nu_{th}=2.52$ (2),
$q=8,\;\nu_{th}=2.6$ (3), $q=4,\;\nu_{th}=2.95$ (4),
without threshold function (5),
$\alpha\tau_{QSO}=2.4\times 10^{5}$ years,
($\nu_{th}$ for given $q$ was obtained from eq. 11).
The rest is the same as in Fig. 1.
}
\end{figure*}

As we can see any model does not explain the
observable $n^{obs}_{QSO}(z)$ at high $z$.
Therefore we have repeated the same  procedure without the equation
for concentration of bright galaxies (11).
Results are presented in the Table 3 and Fig.5.

\begin{table*}[tbh]
\caption{
Solutions $\alpha^{'}\tau_{QSO}$, $\nu_{th}$, $q$ of system of equations (10)
and ratio $K$ of number density of
peaks which has passed through quasar stage to one of bright galaxies.
}
\vspace{0.5cm}
\begin{center}
\begin{tabular}{|c|c|c|c|c|}
\hline
&\multicolumn{3}{|c|}{CDM}&$H\;+\;CDM$ \\
\hline
$n$           & 1            &   0.8        &0.7           &1             \\
\hline
$\alpha^{'}\tau_{QSO}$(years)& $5.73\times 10^{4}$ & $1.83\times10^{7}$ & $6.91\times10^{7} $ & $1.64\times10^{8}$\\
$\Delta t$ (years) & $1.3\times10^{9}$ & $5.19\times10^{8}$ &   0.00              &  0.00         \\
$\nu_{th}$         &   1.09             &   7.26             &   5.89              &  4.95        \\
$q$                &   4.96             &   4.85             &   8.08              &  12.32        \\
$K$                &   7.2              &   0.02             &   0.006             &  0.003        \\
\hline
\end{tabular}
\end{center}
\end{table*}

In this case the parameter $\alpha$ has other interpretation: it is the
fraction of peaks which are selected by threshold function
$p(\nu;\nu_{th}^{QSO},q^{QSO})$ and which pass through quasar stage and
are visible
by Earth observer. Let $\alpha '$ denote it. The ratio of the number density
of all peaks selected in a such way to number density of bright galaxies
$$K=\left.
{\int_{0}^{\infty}\;p(\nu;\nu_{th}^{QSO},q^{QSO})\;N_{pk}(\nu)\;d\nu
\over
\int_{0}^{\infty}\;p(\nu;\nu_{th}^{g},q^{g})\;N_{pk}(\nu)\;d\nu}\right.$$
is calculated for each model and  presented in Table 3.
In all models, with the exception of standard CDM, $K\ll 1$ and can be
interpreted as $\alpha_{1}$ above mentioned, that is  fraction of the
peaks-precursors of bright galaxies passed through QSO stage. Thus, in
these models the peaks resulting in bright galaxies through QSO stage are
higher at average than main part of all peaks. The picture is inverse in
standard CDM, where $K\approx 7$, and may mean that peaks of galactic mass
$M\ge 2\times 10^{11}\;h^{-1}M_{\odot}$ survive after QSO stage and become
bright galaxies if their gravitational potential holes are sufficiently deep.
Whether it is valid or not can be proved only by means of elaborate numerical
simulations of the evolution of such peaks in different cosmological models.
As it follows from Fig. 5, the theoretical number density of QSOs at low $z$
is satisfactory in
all models. At high $z$ only CDM model with $n=0.7$ gives the number
density of QSOs sufficiently close to observable one.
But even this model does not explain the observable
number density of QSOs at $z\ge 4$. Is it possible to explain the whole
observable redshift distribution of quasar abundance given by Schmidt et al.
[\cite{sc91}] in such approach, in principle?

\subsection{Finding of phenomenological spectrum}

        For answer to this question we attempted to solve the system of
equations (10)-(11) with respect to unknown values $\sigma_{0}$, $\sigma_{1}$,
$\sigma_{2}$, $\alpha \tau_{QSO}$, $\nu_{th}$ and $q$. The time delay
$\Delta t$ is assumed to be equal to zero. Solutions are obtained in the
same way and are presented in the Table 4. Quasar abundances $n_{QSO}(z)$
for them are shown in the Fig.6. and match the observable data quite well.

\begin{table*}[tbh]
\caption{
Partial solutions $\alpha\tau_{QSO}$, $\nu_{th}$,  $q$, $\sigma_{0}$,
$\sigma_{1}$ and $\sigma_{2}$ of system of equations (10)-(11).
}
\vspace{0.5cm}
\begin{center}
\begin{tabular}{|c|c|c|c|c|c|}
\hline
$\alpha \tau_{QSO}$(years) &$3.20\times 10^{5}$ &$3.51\times 10^{5}$ &$3.53\times 10^{5}$  &$4.41\times 10^{5}$ &$4.0\times 10^{4}$ \\
$\Delta t$ (years)         & 0.00               & 0.00               &   0.00              &  0.00              &  0.00        \\
$\nu_{th}$               &   2.06             &   1.69             &   1.69              &  1.67              &  1.67        \\
$q$                      &   8.71             &   9.28             &   9.28              &  9.31              &  9.31        \\
$\sigma_{0}$             &   2.38             &   2.77             &   2.78              &  2.81              &  2.82        \\
$\sigma_{1}$             &   1.48             &   0.22             &   0.17              &  0.02              &  0.002       \\
$\sigma_{2}$             &   2.37             &   0.43             &   0.34              &  0.03              &  0.004       \\
\hline
\end{tabular}
\end{center}
\end{table*}

As we can
see, the main feature of all solutions is the relation between momenta:
$\sigma_{1}< \sigma_{2}<\sigma_{0}$. What kind of spectrum have such relations
between its momenta on the galaxy scale ?
We attempted to modify  the standard CDM spectrum  on small scale
by force of reducing and enhancing of its power in order to obtain momenta
of spectra from Table 4. Locations and amplitudes of such modifications were
unknown parameters and were found from 3 equations for given momenta
$\sigma_{1}$, $\sigma_{2}$, $\sigma_{0}$ using MathCad software package.
It was successful for momenta from the column 1 in Table 4 only,
for which parameter  $\gamma \equiv \sigma^{2}_{1}/\sigma_{0}\sigma_{2}=0.3$.
The phenomenological power spectrum of density fluctuations obtained in such
way is
$$P_{ph}(k)=AkT_{CDM}^{2}
\left\{
 \begin{array}{rcl}
&1,& k<k_{0},\\
&e^{1-(k/k_{0})^{p}},& k_{0}\le k \le k_{1},\\
&e^{1-(k_{1}/k_{0})^{p}},& k>k_{1},\\
 \end{array}
\right.
$$
where $k_{0}=0.48h\,Mpc^{-1}$, $k_{1}=0.82h\,Mpc^{-1}$, $p=2.3$, $A$ is the same
as in standard CDM. The spectrum has shelf-like reducing of power at
$k\sim 0.7h\,Mpc^{-1}$ in comparison
with standard CDM normalised to the COBE $10^{o}$ angular scale temperature
anisotropy of CMB ($P_{1\;ph}(k)$ spectrum in Fig. 7).

The correlation
function of galaxies for it does not contradict observable data.
Other cosmological consequences of such spectrum are being investigated yet.

The parameter $\gamma$ for momenta from columns 2-5 in Table 4 is extremely
small (0.04-0.0004), so that phenomenological power spectrum for them,
obviously, must be extraordinary too. Indeed, spectrum with momenta from
column 2 ($P_{2\;ph}(k)$ spectrum in Fig. 7)
has two narrow bumps at $k\approx 0.02$ and $\approx 10h\,Mpc^{-1}$
and shelf-like reducing of power at $k\sim 0.03h\,Mpc^{-1}$ in comparison
with standard CDM. Since its correlation function of bright galaxies
strongly contradicts observable one, the solutions from columns 2-5 in Table 4
can be interpreted as meaningless from physical point of view. So, only
spectra with momenta from column 1 ($\gamma\approx 0.3)$, or close to them,
are cosmologically significant.

\begin{figure}[tbh]
\epsfxsize=8truecm
\centerline{\epsfbox{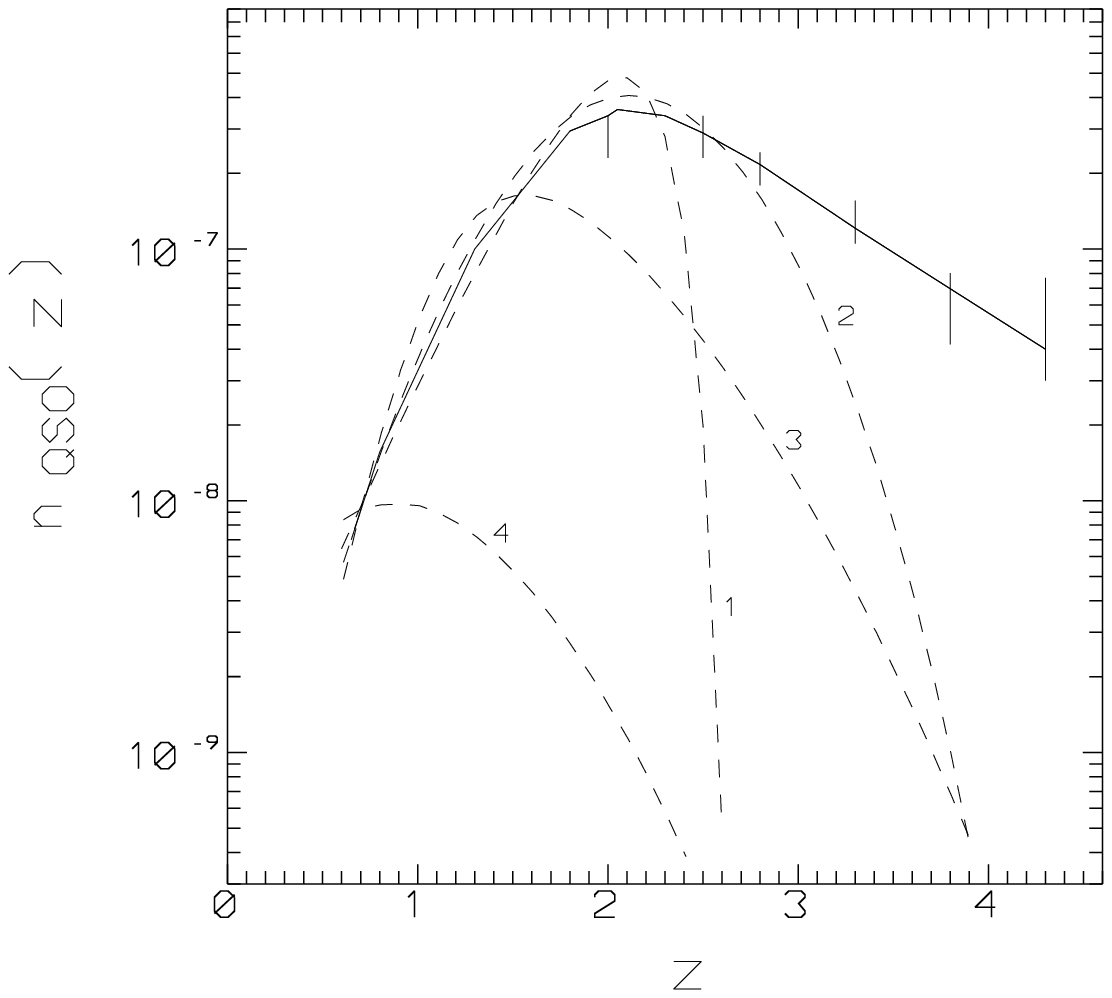}}
\caption{
Quasar abundance against redshift obtained by minimization
of divergence vector including also number density
of bright galaxies at $z \sim 0$:
1 - CDM $n=1$, 2 - CDM $n=0.8$, 3 - CDM $n=0.7$, 4 - H+CDM $n=1$.
}
\end{figure}

\begin{figure}[tbh]
\epsfxsize=8truecm
\centerline{\epsfbox{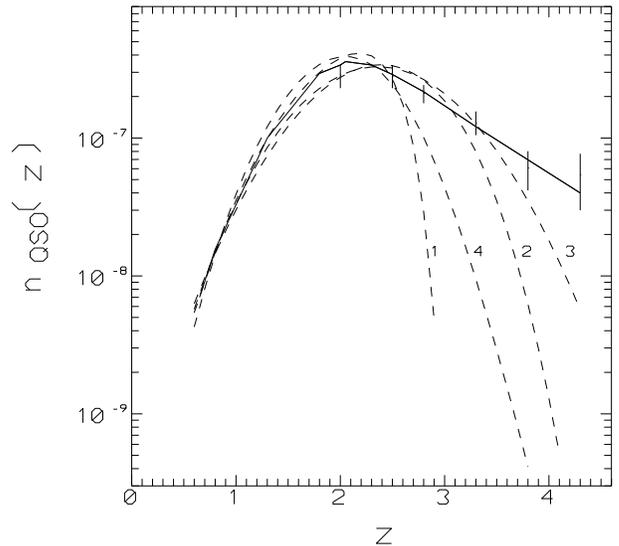}}
\caption{
Quasar abundance against redshift obtained by minimization
of divergence vector without abundance of bright galaxies:
1 - CDM $n=1$, 2 - CDM $n=0.8$, 3 - CDM $n=0.7$, 4 - H+CDM $n=1$.
}
\end{figure}
\begin{figure}[tbh]
\epsfxsize=8truecm
\centerline{\epsfbox{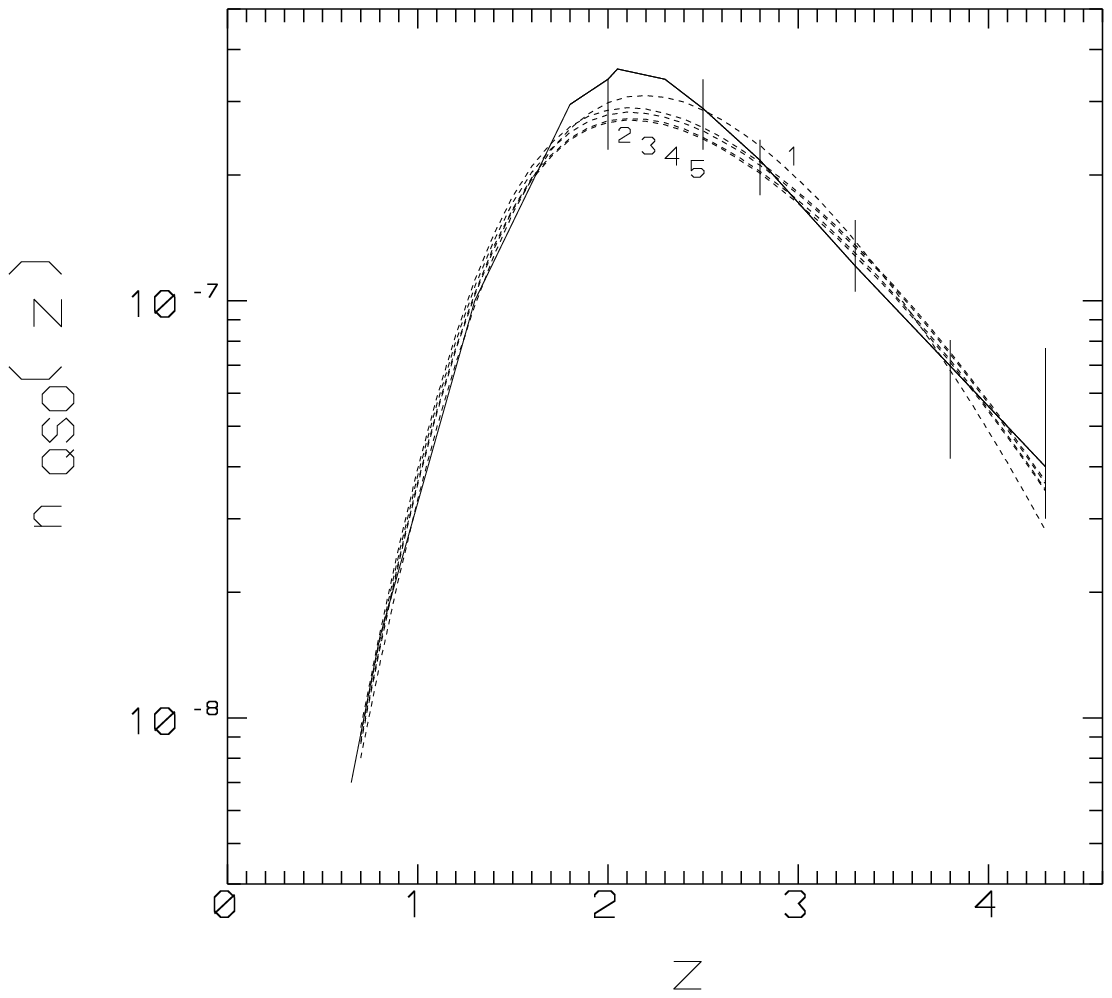}}
\caption{
Quasar abundance against redshift for found momenta of spectra,
parameters of threshold function and $\alpha\tau_{QSO}$.
}
\end{figure}

\subsection{Estimation of fraction of host galaxies}

 Masses of galaxies which pass through quasar stage (host galaxies) are
$M_{g}\ge 2\times 10^{11}h^{-1}M_{\odot}$, so that it is coordinated with the
black hole accretion models of quasar phenomenon [\cite{tu91,lo94,ef88,ka91}].
An optical luminosity of
the bright quasars is $L\ge 10^{47}\;h^{-2}\;erg/s$. Their lifetime
multiplied by $\alpha$ in all cases is $\alpha \; \tau_{QSO} \ge
10^{5}$ years, because we denote $\tau_{QSO}= 10^{5} \; {\tau_{5} /
\alpha}$, where $\tau_{5}$ is $\alpha  \; \tau_{QSO}$ in the units of
$10^{5}$. If the fraction of the black hole mass converted into
optical radiation is $\epsilon$, the ratio of the quasar's mass to
that of the host galaxy is $F$, then
$$M_{g}\approx 2\times 10^{5}\; L_{47}\; \tau_{5} / (\alpha\;\epsilon F)\;
h^{-2},$$
that gives
$$\alpha\approx 10^{-3}\;L_{47}\; \tau_{5} / (\epsilon_{0.1}\;F_{0.01})\;h^{-1,}$$
 where $\epsilon_{0.1} \equiv {\epsilon / 0.1}$, $F_{0.01} \equiv {F /
0.01}$, $L_{47} \equiv {L / 10^{47}}$.

Thus, in the models with phenomenological power spectrum and spectra tested
here only less than $1\%$ of massive galaxies pass through quasar
stage and are visible by Earth observer.

\section{Discussion}

 Recently we [\cite{cn96,nc96}] carried out the similar
calculations of evolution of quasar number density for the same
spectra (CDM with n=1, 0.8, 0.7 and H+CDM) normalised to
$\sigma_{8}=b_{g}^{-1}$, where galactic biasing parameter $b_{g}$
was calculated in the framework of Gaussian statistics of density
peaks [\cite{b86,hn91}] and was equal to
1.40, 1.47, 1.53 and 1.76 respectively. From comparison of results of
this work with previous ones it follows that the dependence of quasar
number density on $z$ is very sensitive to the normalisation of
spectrum in all models. In our previous papers the modified on small scale
standard CDM spectrum for reproducing of the observable data on number
density of QSOs at different redshifts has reducing of power on scale
$1 \le k \le 10h Mpc^{-1}$ and bump on scale $k\sim 10 h\,Mpc^{-1}$
(dotted line in Fig.7).
 Its difference from
spectrum shown in Fig.7 is caused by different power of
initial CDM spectrum on $k \le 1h\,Mpc^{-1}$ as a result of different
normalisation.

It should be remarked here, though formally this method is sensitive to
spectrum on small scale up to $k\sim 10h\,Mpc^{-1}$ when it is smoothed by
Gaussian filter function with $R_{f}=0.35h^{-1}Mpc$ (or top-hat
$R_{TH}=0.6h^{-1}Mpc$) nevertheless we are not sure that calculation of
peak number density of galaxy
scale is exact for cases of spectrum like standard CDM and, especially,
CDM + short wave bump models. This is concerned mainly with the well known
problem 'cloud-in-cloud' unresolved by now in the frame of peak formalizm
used here.

        Now let us compare our results with those by Nusser and Silk 
[\cite{nu93}].
The curve $n_{QSO}(z)$ for $\delta_{c}^{o}=1.33$ (line 6 in Fig.3)
without threshold function, recalculated to
$\alpha \tau_{QSO}=10^{8}$ years, coordinates with
the corresponding curve in Fig.2 of work by Nusser \& Silk 
[\cite{nu93}] quite
satisfactory at high z. The some divergence of these curves at low z are
caused mainly by diferent ways of definition the mass of peaks.

\begin{figure}[tbh]
\epsfxsize=8truecm
\epsfbox{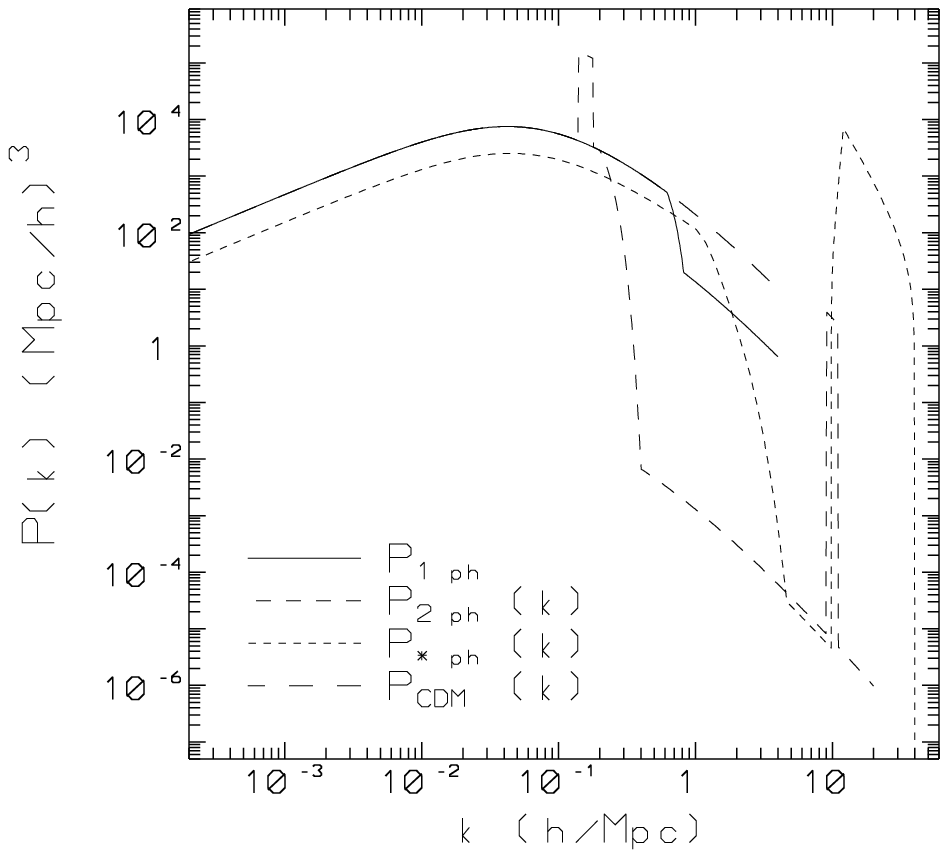}
\caption{
The power spectrum of density fluctuation normalised to product
COBE $\Delta T/T$ at $10^{o}$ angular scale for standard CDM,
its modification for momenta from column 1 in Table 4 ($P_{1\;ph}$),
the same for momenta from column 2 in Table 4 ($P_{2\;ph}$) and
the phenomenological power spectrum with $\sigma_{0}=2.37$, $\sigma_{1}=1.16$,
$\sigma_{2}=1.89$ ($\tau_{QSO}=4.4\times 10^{5}\;years$, $b=1.72$) obtained
by modification of standard CDM normalised to
$\sigma_{8}=b_{g}^{-1}$ ($P_{*\;ph}$).
}
\end{figure}

In our work the value $\delta_{c}^{o}=1.68$ corresponding to
collapse of spherical-symmetrical peaks, is adopted.
Therefore, in our approach the QSOs are formed in spheroidal peaks
only. The fraction of spheroidal peaks is $\le 4\%$ [\cite{dor70}],
the fraction of massive galaxies with $M\ge 2\times 10^{11}\;h^{-1}
M_{\odot}$ passed through quasar stage in all cases does not
exceed $1\%$. So, the
assumption that QSOs appear only in spherical peaks of density
fluctuations is not contradictory in our approach. But its conformity
to reality should be proved or objected by elaborated numerical
simulations of the evolution of QSOs which start from initial
conditions determined at the linear stage of the evolution of peaks.
Substantiations of the value $\delta _{c}^{o}$ as well as dependence or
independence of both quasar lifetime $\tau_{QSO}$ and time delay
$\Delta t$ on redshift $z$ will be the subject of the future works.
c
We put constraints on the cosmological models based on their
ability to reproduce the rise of quasar abundance up to redshifts
$\approx 2.2$ and following fall up to $\approx 4.7$  assuming that
quasar nature is the same at different redshifts and $\tau_{QSO}$,
$\Delta t$ and $\alpha$ do not depend on z.  Moreover,
if we have only 3 unknown values $\alpha \tau_{QSO}$, $\nu_{th}$
and $q$ in equation system (10-11) then we can not reproduce the
$n^{obs}_{QSO}(z)$ given by Boyle et al. [\cite{bo88}]
 and Schmidt et al. [\cite{sc91}]
 with sufficient accuracy in the models tested here.
But when $\sigma_{0}$, $\sigma_{1}$
and $\sigma_{3}$ are supposed to be unknown values too, such reproducing
is possible. Apparently, for other models of quasars,
which explain the rise and fall of the  $n^{obs}_{QSO}(z)$ by
astrophysics of quasars (e.g. [\cite{ha93}]), the constraints on the
cosmological models may be other. But it is the subject of
separate work. Here we only confirm that such dependence of quasar
abundance on z can be explained in the framework of most simple
astrophysical models of quasars, which look rather plausible and
motivated in order to be put in the base of similar research.

The phenomenological power spectrum obtained here ($\S 3.2$) has
shelf-like reducing of power at $k\sim 0.7h\,Mpc^{-1}$. Such spectra
with broken scale invariance is generated in some inflationary
scenarios (e.g. [\cite{am90,got91,st92,pet94}]). The typical
scale and the height of the shelf-like reducing  determined here
can be connected with some constants, which characterise the
underlying inflationary models.

      If we treat the used observational data
more critically and suppose  that $n^{obs}_{QSO}(z)$ given by
Schmidt et al. [\cite{sc91}] and Boyle et al. [\cite{bo88}] is only lower limit
for number
density of QSOs at all $z$ then the next constraints follow from the
presented results: on the small scale the initial power spectrum of
density fluctuations is nearly CDM with $n\ge 0.7$, $b_{g}\le 2.3$
and $\alpha \tau_{QSO}\ge 5\times 10^{5}$ years,
$\Delta t \le 5\times 10^{8}$ years.

\section{Conclusions}

        Thus, the main assumption that quasar phenomenon is active short
term stage of evolution of some small fraction of massive galaxies
with $M\ge 2\times 10^{11}\;h^{-1} M_{\odot}$ and appears in peaks of random
Gaussian density fluctuation field allows to explain the general
feature of $n^{obs}_{QSO}(z)$ given by Boyle et al. [\cite{bo88}]
and Schmidt et al. [\cite{sc91}]: fast increasing of number density from
$z\approx 0$ to $z\sim 2.2$ up to value $\sim 3\times 10^{-7}\;h^{3}
Mpc^{-3}$ and following monotonous decreasing on $z>2.5$.
Apparently, a constant quasar abundance  from $z\approx 2$ to $4$
[\cite{ir91}] can not be explained under these assumptions.

    If the nature of bright QSOs is the same at different redshifts
($\tau_{QSO}$, $\Delta t$, $\alpha$ are constants)
then any model spectrum tested here such
as standard CDM spectrum ($\Omega_{CDM}=0.9$, $\Omega_{b}=0.1$),
tilted one with $n=0.7, 0.8$ and hybrid H+CDM ($\Omega_{CDM}=0.6$,
$\Omega_{HDM}=0.3$, $\Omega_{b}=0.1$, $n=1$) explains the number density
of such QSOs in redshift range $z\approx 0.5-2.2$, but in range $z\approx
0.5-3.5$ only tilted CDM with $n=0.7$ ($\tau_{QSO}\approx 7\times 10^{7}$ years,
$\alpha_{1}\approx 0.01$, $\Delta t=0$) does explain, and in range $z\approx 0.5-4.7$ -
none of them. Consequently more crucial for testing of spectra is number density
of QSOs on high $z$.

        Exact reproducing of number density of QSOs
at $z=0.5-4.7$ under above mentioned assumptions can
 be done in the
flat cosmological model with the density fluctuation power spectrum
which gives  $\sigma_{0}\approx 2.4-2.8$ and $\gamma \le 0.3$
at galaxy scale ($R_{f}=0.35\;h^{-1} \;Mpc$).
It predicts the QSO's lifetime
$\tau_{QSO}\approx(3.5-4.5)\times 10^5 /\alpha$ years, where $\alpha$ is
the fraction of massive galaxies passing through quasar stage. If
physically motivated quasar lifetime is $\sim 10^{7} - 10^{8}$ years,
then $\alpha \approx 0.01-0.001$ that matches the estimations of $\alpha$
done above.

        Thus, redshift distribution of quasar abundance is very sensitive
to the amplitude and slope of the initial power spectrum of density
perturbations on galaxy scale. But it can be effective test when
complete number density of QSOs with mass larger than fixed value will be
known with confidence for all redshifts and physical parameters
connected with nature of quasar phenomenon (the quasar lifetime
$\tau_{QSO}$, time delay $\Delta t$, fraction of galaxies which pass
through quasar stage $\alpha$, linear amplitude of peaks $\delta_{c}^{o}$
collapsing just now and threshold function) will be substantiated.
The last needs the detailed investigation
of connection of quasar phenomenon with initial configuration of
density fluctuation by the help of numerical simulations of its
evolution.

We acknowledge useful discussions with B. Hnatyk, V.N. Lukash and
B.V. Komberg.

\end{document}